\documentclass[12pt]{iopart}

\usepackage{graphicx}
\usepackage{multirow}
\usepackage{color}
\usepackage{colortbl}

\begin{document}

\title[]{Bounded link prediction for very large  networks}

\author{ Wei Cui, Cunlai Pu, Zhongqi Xu}
\address{School of Computer Science and Engineering, Nanjing University of Science and Technology, Nanjing 210094,  China
}
\ead{pucunlai@njust.edu.cn}
\begin{abstract}
Evaluation of link prediction methods is a hard task in very large complex networks because of the inhibitive computational cost. By setting a lower bound of the number of common neighbors (CN), we propose a new framework to efficiently and precisely evaluate the performances of CN-based similarity indices in link prediction for very large heterogeneous networks. Specifically, we propose a fast algorithm based on the parallel computing scheme to obtain all the node pairs with CN values larger than the lower bound. Furthermore, we propose a new measurement, called self-predictability, to quantify the performance of the CN-based similarity indices in link prediction, which on the other side can indicate the link predictability of a network.
\end{abstract}

\maketitle
\section{Introduction}
The scale and complexity of real-world systems such as social systems grow unprecedentedly, which makes the prediction of such systems more and more challenging, and on the other hand attracts more and more attention from both industry and research \cite{clauset08,Newman09,wangwx11,Dorogo,Boccaletti,lu15}. Link prediction problem is described as quantifying the likelihood of yet unknown associations between individuals in networks. Generally, there are two kinds of link prediction problems \cite{linben,lu11}. The first one is predicting the future links given the current topology of a network. The second one is inferring the missing links that are likely to exist in a static snapshot of a network. Link prediction has wide applications in social systems \cite{HASAN,Wangd11}, biological systems \cite{huynen,Barzel13}, scientific systems \cite{wang13,Petersen}, etc. For example, in recommender systems \cite{tang13,lv12}, they could either suggest people (items) whom you might find interesting enough, or the friends (items) you have already known, but just not yet connected online. In national safety applications, link prediction can help to identify the hidden terrorist groups or criminal relations \cite{latora04,krebs02}. In bioinformatics, link prediction is employed to find the interactions between proteins \cite{sharan07} or the side effects of drugs \cite{Lin13}. In scientific research, link prediction can be used to find the experts \cite{Pavlov} or future potential coauthors \cite{sun11,benchettara} based on the current co-authorship networks. In all these applications, link prediction facilitates the evolution or enhances the integrality of related systems.

Link prediction methods may be broadly divided into three groups \cite{lu11} : maximum likelihood algorithms, probabilistic models, and similarity-based strategies. Compared with the first two types of approaches, similarity-based strategies seem more promising, because of their simplicity and relatively lower computational cost. In the similarity-based methods, unconnected node pairs are assigned the similarity scores, and the node pairs with high similarity scores are assumed to be linked by edges with high probability. However, it is hard to define node similarity indices, partly because node attributes are not easy to be obtained. Thus, many researchers aim to propose similarity indices only with the knowledge of network structure, which can be further categorized into three types according to the amount of information used in the similarity computation: local indices \cite{newman01,adamic03,zhout,tsorensen}, global indices  \cite{katz,leicht,jeh,chebotarey}, and quasi-local indices \cite{lu09,wliu,papadimitriou}. Local indices require the information of the local structure of nodes to determine the similarity of nodes. The first and most widely studied local index is the common neighbors (CN) index \cite{linben}, which quantifies the similarity of a pair of nodes as the number of neighbors they have in common. Many other local indices are the extensions of the CN index \cite{lu11}.

Although local indices are promising in link prediction, they are hard to be evaluated in large real-world networks. First, the current evaluation framework require to calculate the similarities of a large number of  node pairs with the worst time complexity of O($|V|^3$), where $V$ is the node set \cite{lu11}.  Even if we use the multi-core cluster in the computation, the cost only decreases at most linearly. On the other hand, the current metrics such as the precision \cite{Geisser}, AUC \cite{Hanley}, etc. for evaluating the performance of similarity indices have limitations when applied to large-scale networks.  The number of edges in the complement graph of a large real-world network is usually much larger than that in the original network, which makes the value of precision tend to zero, and leads to a large variance of AUC. Recently, several fast link-prediction algorithms \cite{papadimitriou,sschelter} based on MapReduce computational model \cite{jdean} were proposed in the literature. These algorithms are demonstrated to work efficiently in the calculation of CN-based similarity indices, since they delicately divide the computation into the ¡°map¡± phase and the ¡°reduce¡± phase, and the computation may be parallelized into clusters of many machines. However, these fast algorithms do not reduce the time complexity essentially, and they are still not efficient for large dense networks with millions of nodes.

In fact, most of the real-world networks have heterogeneous topological structures, and  link recommendation usually happens in the relatively dense areas of the networks. For example, we are more interested in individuals which have a large number of friends, and we would like to recommend these ``hot" individuals to the others. Based on these facts, we propose a bounded link prediction framework, in which we set a lower bound of CN values. With the lower bound, we propose a fast parallel algorithm for calculating CN values based on the MapReduce model. Also, we propose a new metric for evaluating the performance of the CN-based indices. With our fast algorithm and the new measurement, we can efficiently and precisely evaluate the CN-based indices in link prediction for very large real-world networks.

\section{Traditional algorithms based on MapReduce}

In this section, we discuss the traditional MapReduce based algorithms \cite{papadimitriou,sschelter} for calculating the CN values  and discuss their limitations from the perspective of computational complexity.

MapReduce \cite{jdean} is a widely used programming model for dealing with searching, sorting, and many other tasks related to large-scale datasets. Programmers find the MapReduce system easy to use in that it automatically parallelizes the tasks across large-scale clusters of machines, handles machine failures and schedules inter-machine communications. User only needs to specify the computation task in terms of a map and a reduce function. The map function takes an input key/value pair and produces a set of intermediate key/value pairs. Then, all the intermediate pairs are grouped by the key and passed to the reduce function. In the reduce function, the values for a key are merged together to form a smaller set of values. For the traditional pair generating algorithm \cite{papadimitriou}, node indices and node adjacencies are specified as the key/value inputs of the map function.
In the phase of ``map", neighbors of a node are paired with each other, and each node pair is taken as the key and  assigned a value (score) ``1".  In the phase of ``reduce", the values for each key are summed which are the desired CN values.

This algorithm can be further improved by means of vectorization \cite{sschelter}. In the vectorization algorithm, the value of a key is set to be an accompanied group (see Fig. 1 for the illustration of accompanied groups). Thus, the times of data transmission from the map function to the reduce function are greatly reduced.
For example, assume there are several node pairs incident with node $v7$ which are $(v1, v7)$, $(v2, v7)$, $(v4, v7)$, $(v5, v7)$. For the pair generating algorithm, these node pairs are all different keys and will be sent to the reduce function one by one. However, in the vectorization algorithm, these node pairs are transformed into a key/value pair, where the key is node $v7$ and the related value is $v7$'s accompanied group $\{v1, v2, v4, v5\}$, and then the intermediate key/value pair are emitted to the reduce function.
  The pseudocodes for the pair generating algorithm and the vectorization algorithm are in \textbf {Appendix A}.

We present a small example to further illustrate the above two algorithms. As shown in Fig. 1, the given network contains 8 nodes and 15 edges. First, we get the node adjacencies based on the given network. Then, with the node adjacencies as the inputs, we  get all the non-zero CN values for the given network by using the pair generating algorithm and the vectorization algorithm respectively.

\begin{figure}
 \centering
%\onefigure{F3.eps}
\includegraphics[width=\textwidth,height=8in]{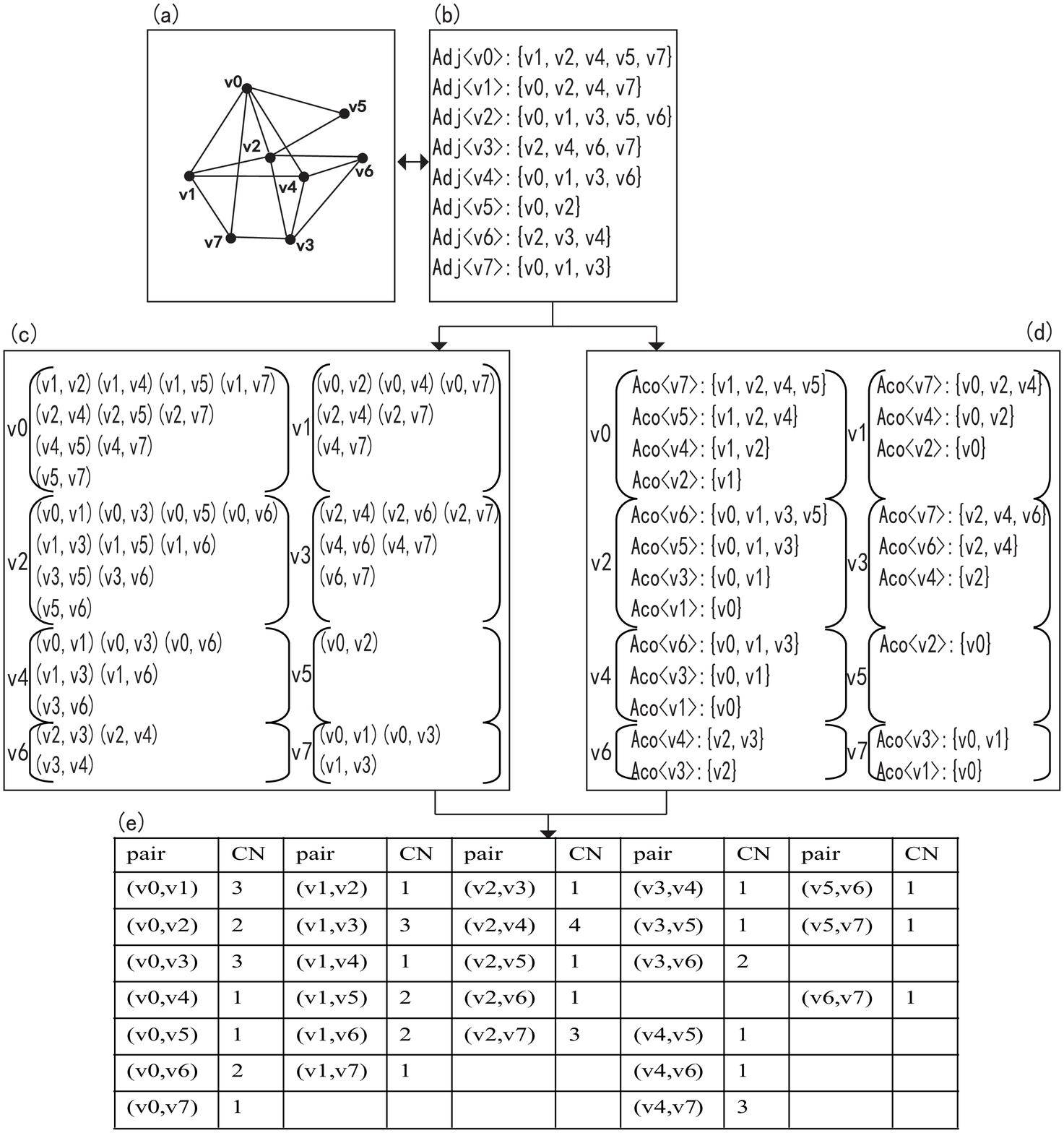}
\caption{Illustration of the procedure for calculating the CN values with the pair generating algorithm and the vectorization algorithm. (a) is the given network. (b) shows the  node adjacencies of the network. For instance, ``Adj$\langle v0 \rangle$" represents the adjacencies of node $v0$, which is ``$\{v1,v2,v4,v5,v7\}$".  (c) and (d) are the  outputs of the map function for the pair generating algorithm and the vectorization algorithm respectively. For instance, in the first line of (c), ``$(v1, v2)(v1, v4)(v1, v5)(v1, v7)$" represents the node pairs incident with node $v1$,  which are generated based on Adj$\langle v0 \rangle$.  In the first line of (d),
``Aco$\langle v7 \rangle$" represents the accompanied group of node $v7$, which is $\{v1, v2, v4, v5 \}$.
 Assuming the adjacencies of a node  is $\{s_0, s_1, s_2, \cdots\}$, the accompanied group of node $s_i$ is defined as  $\{s_{0}, s_{1}, \cdots ,s_{i-1}\}$ which is the node set that contains all the nodes  ordered in front of node $s_i$.
(e) shows the results of the reduce function which are  same for  both of the two algorithms.  }
%\label{fig.1}
\end{figure}

Compared to the other non-parallel algorithms, the  MapReduce based algorithms are tested to be very efficient in the calculation of  CN-based similarity indices.  The reason is that only the node adjacency information is needed in the calculation of CN values, and thus these  MapReduce based algorithms  can easily parallelize the computation in clusters of machines.
Assume a graph $G(V,E)$, where $V$ is the node set and $E$ is the edge set. The average degree is $\langle k \rangle =2|E|/|V|$. In the best case, all the nodes have the same node degree $\langle k \rangle$. Then,  the number of node pairs generated based on the adjacency of a node is $C(\langle k \rangle, 2)=\langle k \rangle (\langle k \rangle-1)/2$, and the total number of generated node pairs is $\langle k \rangle (\langle k \rangle-1)|V|/2$. Thus, the time complexity in the best case is O$(|E|*\langle k \rangle)$. Actually, in a sparse network, if all the degrees of nodes are less than a constant value, the time complexity is close to O$(|V|)$.  The time complexity increases with the heterogeneity of degree distribution. This can be illustrated by the following inequation:
\begin{equation}
C(\langle k \rangle, 2)+ C(\langle k \rangle, 2)< C(\langle k \rangle+\Delta, 2)+ C(\langle k \rangle-\Delta, 2), \Delta>0,
\end{equation}
which always holds.
Under the parallel computing environment, the time complexity  decreases by a constant factor depending on the number of machines. Compared to the pair generating algorithm, the vectorization algorithm only reduces the transmission times from O$(|V|^3)$ to O$(|V|^2)$ (in the worst case), while the total numbers of generated node pairs for the two algorithms are the same. 
Thus, the time complexity of the vectorization algorithm is the same as that of the
pair generating algorithm, and it is faster than the pair generating algorithm only by a
constant factor.
Based on the above analysis, we obtain that the large time complexity of the CN-based link prediction methods results from the generation of large number of node pairs. In the worst case, if every node pair do have common neighbors, the lower bound of the ordinary complexity for any algorithm is $\Omega(|V|^2)$.

\section{Our algorithm}
In this part, we firstly introduce the idea of our algorithm which is originated from the scale-free properties of real-world networks. Then, we present two lemmas which is related to the filtering operation in our algorithm. Finally, we introduce the four steps of our fast algorithm.
\subsection{The idea}
Most of real-world networks have heterogeneous topological structures, and present the scale-free property  \cite{reka02}. In a scale-free network, a relatively small fraction of nodes called hubs have a large number of neighbors, while most of the other nodes just have a small number of neighbors. The node degree distribution of a scale-free network obeys the power-law. The imbalance of node degrees becomes more serious with the evolution of the networks according to the preferential attachment rule \cite{barabasi99}. The heterogeneity of the topological structures of many real-world networks is further magnified by the distributions of the CN values.    Fig. 2 presents the simulation results of the CN distributions for six large-scale real-world networks. Clearly, we see that for all the six networks most of the node pairs just have small CN values, while a relatively small number of node pairs have large CN values. Generally, nodes incident with the node pairs of small CN values may  have small  degrees, and these small-degree nodes form the sparse area of a network. However, nodes incident with the node pairs of large CN values should have large node degrees, and they constitute the dense area of a network.
Usually, the dense area better embodies the organization rule of a network than the sparse area. On the other hand, in real situation we are more likely to recommend hot individuals to the others, and the probability of interaction between hot individuals is much larger than that between inactive individuals (which just have few connections with others). Based on these facts, we introduce a lower bound $L$ of CN values in link prediction to filter in advance the node pairs which originally have very small chance to be connected by edges. Since in most real-world networks, a large fraction of node pairs  have small  CN values, plenty of node pairs of CN values no greater than $L$ will be filtered,  which makes the computational cost greatly reduced.

\begin{figure}
 \centering
%\onefigure{F3.eps}
\includegraphics[width=\textwidth,height=3.5in]{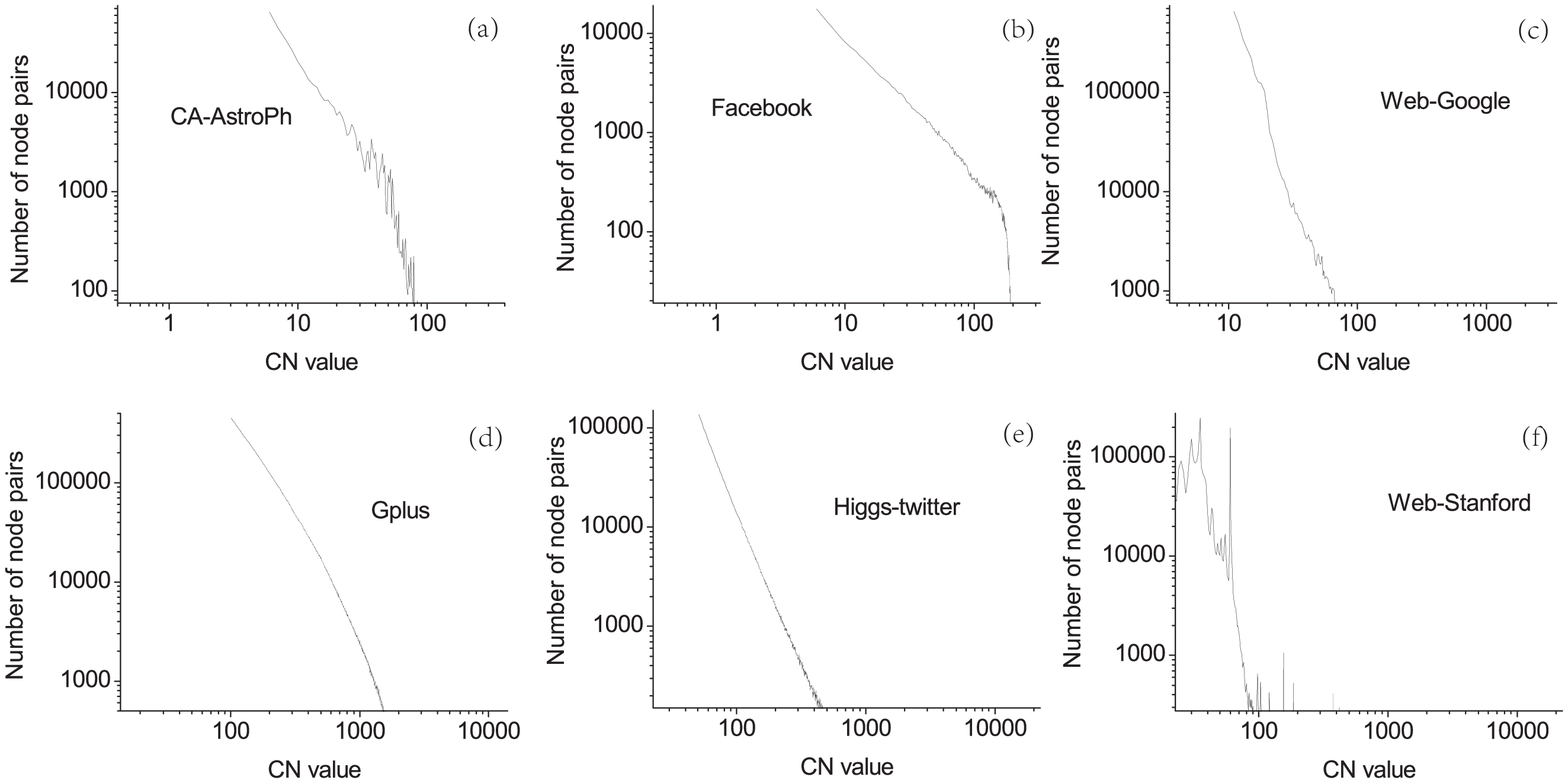}
\caption{Log-log plot of CN distributions for six large-scale real-world networks. The CN values of all the node pairs, either connected by edges or not, are  computed. The statistics of these networks are presented in Table 2. }
%\label{fig.2}
\end{figure}

\subsection{Lemmas}
In our algorithm, we filter the node pairs with CN values  no greater than $L$ in advance  based on  two lemmas, which are as follows:

\textbf{Lemma 1}. In a network, if the number of neighbors of a node is no greater than $L$, we can simply
 ignore the node pairs that contain this node, since these node pairs can not
have more than $L$ common neighbors.

In the implementation, we  filter the node adjacencies of those nodes which have no greater than $L$ neighbors. Note that we can not remove those nodes from the other  node adjacencies either, since those nodes may be the common neighbors of the other node pairs.

 \textbf{Lemma 2}.  In the remaining network (after filtering the original network based on \textbf{Lemma 1}), if a node  appears at most in $L$ node adjacencies,  this node can not be in the desired node pairs.
 \textbf {Lemma 2} is the inverse presentation of \textbf{ Lemma 1}.

 \textbf{Proof of Lemma 2}. Let's assume that a node $u$  is in the node adjacencies. Then $u$ has at
least one accompanied group (Based on our definition of the accompanied group,  the first node in the node adjacencies has no accompanied group. However, we can simply modify the order of the nodes in the node adjacencies to ensure that $u$ has an accompanied group.)  For instance, in Fig. 1(d) node $v7$ has 3 accompanied groups marked with Aco$\langle v7 \rangle$.
Also, in an accompanied group of node $u$ every node is unique, while for several accompanied
groups of node $u$, there may be overlapping nodes. For instance, in Fig. 1(d) node $v1$ appears in two accompanied groups of node $v5$.  For an accompanied group of node $u$, every
element $i$ will be used to generate a node pair $(i, u)$ with score 1. Thus, the total score
(or CN value) of node pair $(i, u)$ is definitely no greater than the number of accompanied
groups that node $u$ has. Therefore, if node $u$ appears at most in $L$ node adjacencies (which means $u$ has no greater than $L$ accompanied groups),  the CN value
of any node pair $(*, u)$ will be no greater than $L$. Then, these node pairs $(*, u)$ can be filtered
 in advance.
\subsection{Steps of our algorithm}
Our fast algorithm is an improved version of the traditional MapReduce-based algorithms \cite{papadimitriou,sschelter} in that it filters the irrelevant node pairs with the above two lemmas, and focuses on dealing with the node pairs of CN $> L$ under the MapReduce scheme. Specifically, our fast algorithm can be divided into the following four steps.

\textbf{Step 1}: we filter the original network based on \textbf{Lemma 1}. Then,  the node adjacencies of size larger than $L$ are reserved.
The filtering of the node adjacencies makes the original undirected network become a directed network, as shown in Fig. 3.

\begin{figure}
 \centering
%\onefigure{F3.eps}
\includegraphics[width=\textwidth,height=2.5in]{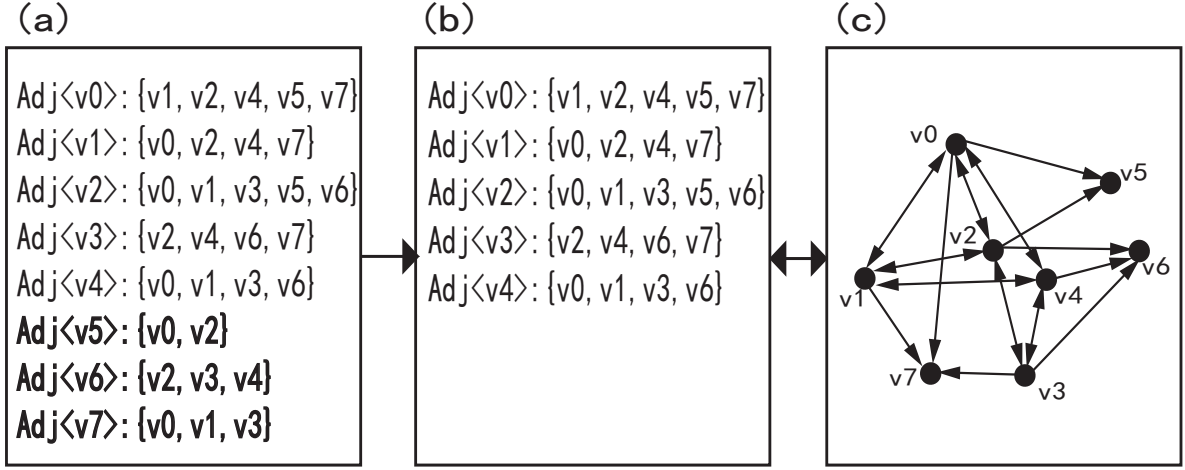}
\caption{Illustration of the first step of our algorithm. The given network is the same as in Fig. 1(a). The lower bound of CN is $L=3$.
We see that node adjacencies Adj$\langle v5 \rangle$, Adj$\langle v6 \rangle$, and Adj$\langle v7 \rangle$ highlighted in bold are filtered. The residual node adjacencies (b)  can be visualized by a directed network  (c).}
%\label{fig.3}
\end{figure}

\textbf{Step 2}: we change the directions of all the edges in the directed network, and generate the new node adjacencies based on the modified directed network, as shown in Fig. 4.

\begin{figure}
 \centering
%\onefigure{F3.eps}
\includegraphics[width=5in,height=2.5in]{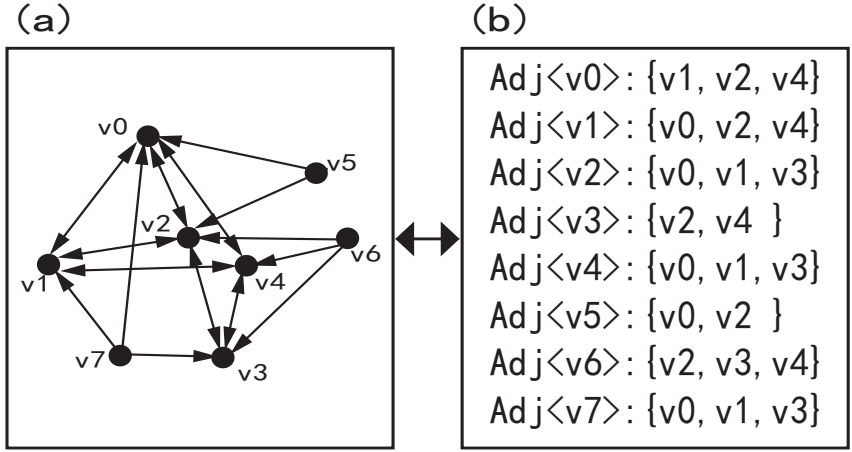}
\caption{Illustration of the second step of our algorithm. (a) is the new directed network generated by changing the directions of all the edges of the directed network in Fig. 3(c).
(b) is the corresponding node adjacencies of (a).  Note that Adj$\langle v5 \rangle$, Adj$\langle v6 \rangle$, and Adj$\langle v7 \rangle$ filtered in \textbf{Step 1} appear again.  }
%\label{fig.4}
\end{figure}

\textbf{Step 3}: based on the new node adjacencies from \textbf{Step 2}, we firstly generate all the accompanied groups. In shared memory environment, we can just emit the addresses and the sizes of accompanied groups instead of the contents of accompanied groups in order to further reduce the amount of data transmission. Then, the intermediate key/value pairs is set to be ``($c, \ll \mbox{Adj}\langle a \rangle, b \gg$)", where  $c$ is a node in Adj$\langle a \rangle$,  and $b$ is the ranking of  $c$ in Adj$\langle a \rangle$,  which is equal to the size of Aco$\langle c \rangle$. Then, we filter the nodes (and their related accompanied groups), which have less than  $L$ accompanied groups according to \textbf{Lemma 2}, as shown in Fig. 5.

\begin{figure}
 \centering
%\onefigure{F5.eps}
\includegraphics[width=\textwidth,height=4in]{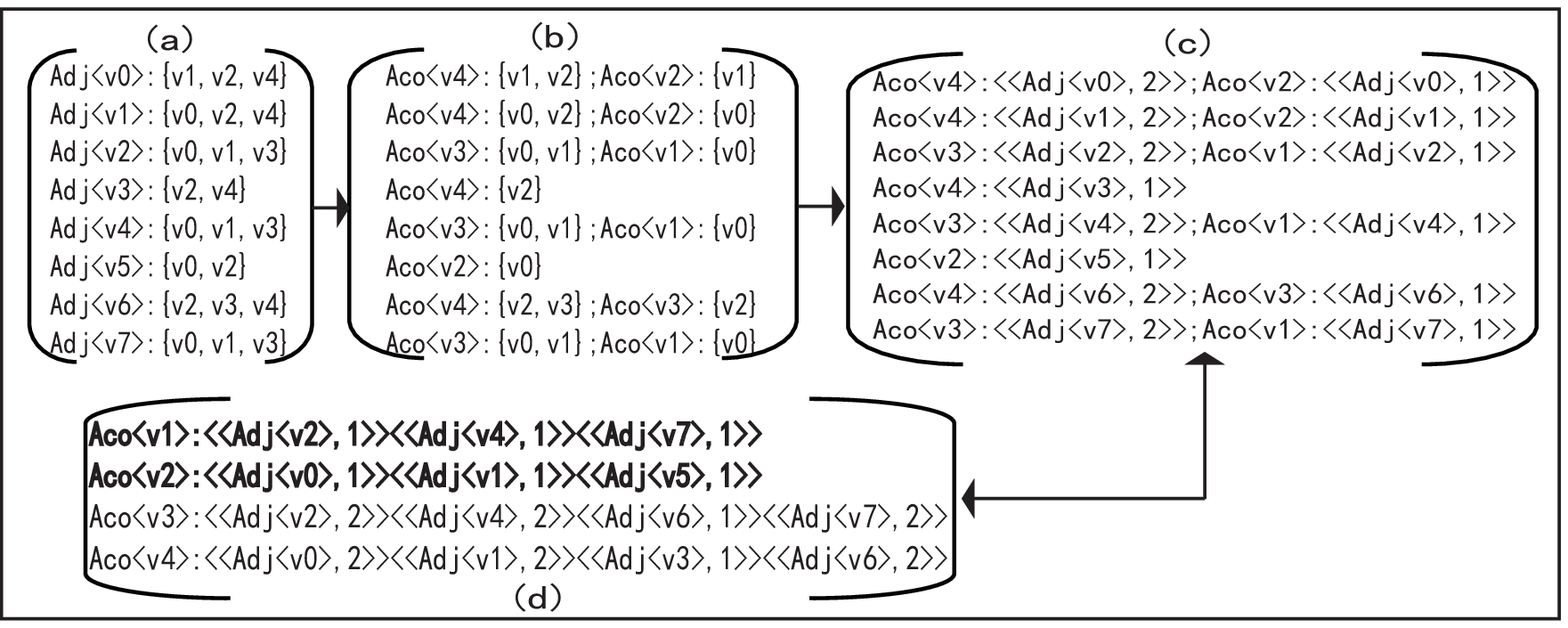}
\caption{Illustration of the third step of our algorithm.
 (a) presents the node adjacencies generated in  \textbf{Step 2}. (b) shows the accompanied groups generated based on (a). (c) gives the ``address and size" representation of accompanied groups. For example, Aco$\langle v4 \rangle$: $ \ll \mbox{Adj}\langle v0 \rangle, 2 \gg$ equals Aco$\langle v4 \rangle$: $\{v1, v2\}$.
  (d) is the ordered version of (c). Aco$\langle v1 \rangle$ and Aco$\langle v2 \rangle$ highlighted in bold are filtered, since both of their numbers are 3 which is no greater than $L$ ($L=3$).
 }
%\label{fig.5}
\end{figure}

\textbf{Step 4}: based on the residual accompanied groups from \textbf{Step 3}, we finally obtain the desired key/values pairs, where key is the node pair, and value is the CN value. Note that although we execute the filter operations in \textbf{Step 1} and \textbf{Step 2} respectively, there might still be some node pairs of CN values no greater than $L$ in the final results, as shown in Fig. 6.

\begin{figure}
 \centering
%\onefigure{F6.eps}
\includegraphics[width=\textwidth,height=2in]{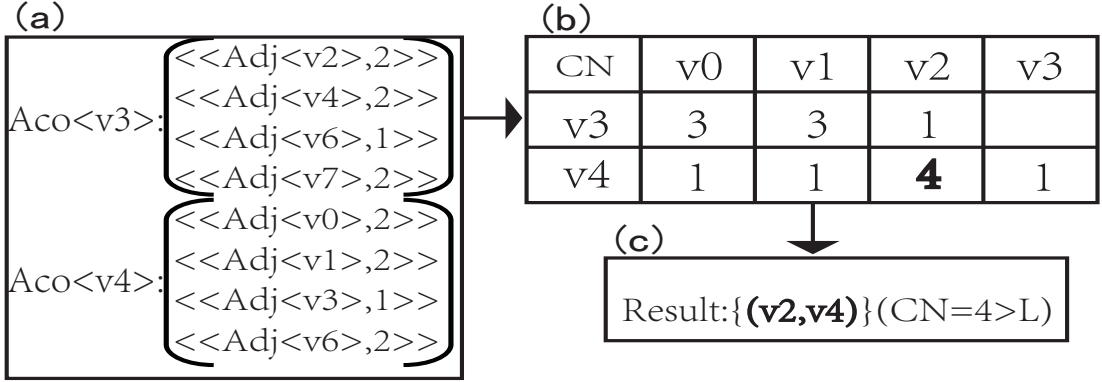}
\caption{Illustration of the fourth step of our algorithm. (a) shows the residual accompanied groups. (b) shows the final generated node pairs and their related CN values. (c) presents the desired results. $L=3$.}

%\label{fig.6}
\end{figure}

The first three steps of our algorithm are very fast. Most of the computational cost lies in \textbf{Step 4}. However, after the twice filtering, the input data of \textbf{Step 4} is greatly reduced. On the other hand, calculations based on the accompanied groups of different nodes are independent, which makes them easy to be parallelized.  The efficiency of our algorithm is also dependent on $L$. Clearly, we have $0\le L<k_{max}$, where $k_{max}$ is the maximum node degree in the network. Large $L$ means that a lot of  node pairs will be filtered, which makes our algorithm very fast. The pseudocode of our algorithm is shown in Table 1.
%%%%%%%%%%%%%%%%%%%%%%%%%%%%%%%%%%%%%%%%%%%%%%%%%%%%%%%%%%%%%%%%%%%%%%%%%%%%%
\begin{table}
\newcommand{\tabincell}[2]{\begin{tabular}{@{}#1@{}}#2\end{tabular}}
\centering
\begin{tabular}{|l|l|}\hline
\tabincell{l}{Step 1} & \tabincell{l}{Step 2}\\\hline
\tabincell{l}{
map($u$, adj$\_$list[$u$]) $\lbrace$
  \\ \quad if (adj$\_$list[$u$].size $>$ $L$)
    \\ \qquad emit($u$, adj$\_$list[$u$])
  \\ \quad else
    \\ \qquad just remove vertices $u$ and \\ adj$\_$list[$u$]
\\ $\rbrace$
} &
\tabincell{l}{
map($u$, adj$\_$list[$u$]) $\lbrace$
\\ \quad  foreach $v$ in adj$\_$list[$u$]
\\ \qquad   emit($v$, $u$);
\\ $\rbrace$
 }
 \\ \hline
 % step 3
\multicolumn{2}{|l|}{Step 3}\\\hline
\multicolumn{2}{|l|}{map($v$, adj$\_$list[$v$]) $\lbrace$ }\\
\multicolumn{2}{|l|}{\quad for $i=1$ to adj$\_$list[v].size()$-1$}\\
\multicolumn{2}{|l|}{\qquad emit(adj$\_$list$[v][i]$, $\ll$ adj$\_$list$[v]$, $i$ $\gg$);}\\
%\multicolumn{2}{|l|}{\quad // $\ll$ addr,len $\gg$ can be O($1$) alias in multi-thread environment}\\
\multicolumn{2}{|l|}{$\rbrace$}\\
\multicolumn{2}{|l|}{reduce($x$,aco$\_$list$[x]$) $\lbrace$ }\\
\multicolumn{2}{|l|}{\quad if(aco$\_$list$[x].$size $>L$ )}\\
\multicolumn{2}{|l|}{\qquad emit($x$, aco$\_$list $[x]$)}\\
\multicolumn{2}{|l|}{$\rbrace$} \\\hline
% step 4
\multicolumn{2}{|l|}{Step 4}\\\hline
\multicolumn{2}{|l|}{map($x$, aco$\_$list[$x$])$\lbrace$}\\
\multicolumn{2}{|l|}{\quad hs $=$ hash$\_$map()}\\
\multicolumn{2}{|l|}{\quad \textbf{foreach} $\ll$ addr,  len $\gg$ in aco$\_$list$[x]$}\\
\multicolumn{2}{|l|}{\qquad for $i=1$ to len$-1$}\\
\multicolumn{2}{|l|}{\qquad \quad hs[($^\ast$addr)[$i$]]$++$}\\
\multicolumn{2}{|l|}{\quad \textbf{foreach}(key, value) in hs}\\
\multicolumn{2}{|l|}{\qquad if(value$>L$)}\\
\multicolumn{2}{|l|}{\qquad \quad emit((key, $x$), value)}\\
\multicolumn{2}{|l|}{$\rbrace$} \\\hline
\end{tabular}
\caption{The pseudocode of our fast algorithm}
\end{table}
%%%%%%%%%%%%%%%%%%%%%%%%%%%%%%%%%%%%%%%%%%%%%%%%%%%%%%%%%%%%%%%%%%%%%%%%%%%%%%%%%%%%%%%%%%%%%%%%5

\begin{figure}
 \centering
%\onefigure{F7.eps}
\includegraphics[width=\textwidth,height=2.5in]{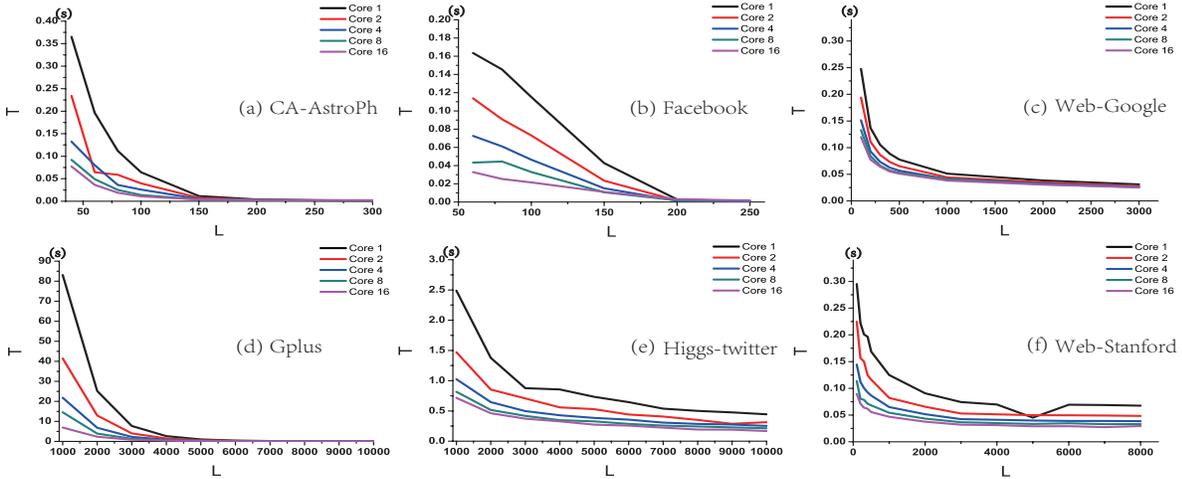}
\caption{ Execution time $T$ vs lower bound $L$ for large-scale real-world networks. Multi-core processors are used in the computation. The core number is increasing from 1 to 16.}
%\label{fig.7}
\end{figure}

\section{Implementation}
In our computing environment, there is a shared-memory machine with multi-core processors running 64-bit Linux with 8GB of memory. Each task is allowed to use 2GB of memory at most.  The time complexity of each of the first three steps in our algorithm is O($|E|$). Thus, the first three steps  can be executed in serial or by employing parallel computing schemes such as  MapReduce or Resilient Distributed Datasets \cite{zaharia} (RDD, see \textbf{Appendix B}). The fourth step of our algorithm has the time complexity of O($|E|*\langle k\rangle$), which is an order of magnitude larger than that of the first three steps. On the other hand, in the fourth step the computation of CN values can be partitioned into many separate running tasks, which make it easy to parallelize the computation in our environment.
We test the performance of our algorithm on six large-scale real-world networks \cite{leskovec-ca-astroph,jmcauley-facebook-gplus,leskov-web-Google-web-stanford,Domenico-higgs-twitter}, of which the statistics are shown in Table 2. We show the results of $L$ vs. execution time $T$ (time for the computation of CN values) in Fig. 7. We see that $T$  is very small, and decreases fast with increase of $L$. For examle,   Gplus  has hundreds of billions of edges, while $T$ of it is just around several tens of seconds. When multi-core processor is used in the computation, $T$  linearly decreases to a few seconds, as is shown in Fig. 7(d).

\begin{table}
\centering
\begin{tabular}{lcc}
\hline
Network&Nodes&Edges \\ \hline
\rowcolor[gray]{.7}
CA-AstroPh\cite{leskovec-ca-astroph}&18,772&198,110\\
\rowcolor[gray]{.9}
Facebook\cite{jmcauley-facebook-gplus}&4,039&88,234 \\
\rowcolor[gray]{.7}
Web-Google\cite{leskov-web-Google-web-stanford}&875,713&5,105,039\\
\rowcolor[gray]{.9}
Gplus\cite{jmcauley-facebook-gplus}&107,614&13,673,453\\
\rowcolor[gray]{.7}
Higgs-twitter\cite{Domenico-higgs-twitter}&456,631&14,855,875\\
\rowcolor[gray]{.9}
Web-Stanford\cite{leskov-web-Google-web-stanford}&281,903&2,312,497\\
\hline \\
\end{tabular}
\caption{The statistics of six large-scale real-world networks.}
\end{table}

\section{Application}
In this part, we firstly discuss the limitations of two widely used metrics in link prediction, which are  precision and AUC. Then we propose a new metric for evaluating the link prediction in very large networks. We name this new metric as self-predictability, since this quantity can reflect the predictability of a network. Finally, we show some simulation results on the self-predictability of several real-world networks.
\subsection{Limitations of traditional metrics}
Link prediction aims to quantify the likelihood of the existence of a link between two disconnected nodes. To evaluate the prediction ability of a similarity index, the original edge set $E$ is usually randomly divided into two parts: a training set $E^T$ and a probe set $E^P$. Clearly, $E^T \cup E^P =E$ and $E^T \cap E^P =\emptyset$. Assume that a universal set $U$ contains all the $|V|(|V|-1)/2$ links. Then, $U-E$ is the nonexistent link set. Each of the links in $U-E^T$ is given a similarity score based on the given similarity index. Precision is defined as the fraction of links, which belong to $E^P$, in the top $r$ links. AUC is defined as the possibility that a link randomly chosen from $E^P$ has a larger similarity value than a link randomly chosen from $U-E$. Usually, we conduct $n$  independent comparisons. Suppose that there are $n'$ times that the link from $E^P$ has a larger similarity value than the link from $U-E$, and $n''$ times that they have the same score, then AUC is calculated as follows:
\begin{equation}
AUC=\frac{n'+0.5n''}{n}.
\end{equation}
For real-world networks, the average node degree $\langle k \rangle$ are usually far smaller than O$|V|$. Thus, when the network size $|V|$ is large, $|U-E|$ is much larger than $|E|$. Then, we get:
\begin{equation}
\frac{|E^P|}{|U-E^T|} \to 0,  \mbox{if}\quad|V| \to \infty.
\end{equation}
This equation indicates that when $|V|$ is large enough, precision should be very small and close to zero.
For AUC, if $|U-E|$ is large, the CN values of links in $U-E$ have large variance. This means that in order to obtain an accurate AUC, the comparison times $n$ should be large enough, which requires a large computational cost.
Therefore, for very large real-world networks, it is inadequate to use precision or AUC to evaluate the performance of similarity indices in  link prediction.
\subsection{Self-predictability}
We propose a new metric, namely self-predictability, to evaluate the performance of a similarity index in link prediction.
The definition of self-predictability $\delta$ is as follows:
\begin{equation}
\delta =\frac{|F(G,L) \bigcap G|}{|F(G,L)|},
\end{equation}
where $G$ is the original graph,  and $L$ is the lower bound of CN values. $F(G,L)$ is the function of $G$ and $L$, which generates the node pairs of CN $>L$. The calculation of self-predictability is much easier than that of  precision and AUC in that it needs not to divide the network into the train and probe sets. Therefore, self-predictability requires lower computational complexity than  precision and AUC. Also, self-predictability reflects to which extent  a network can be predicted by the CN index, and on the other hand indicates prediction precision of the CN index.  $\delta=1$ means that if two nodes have more than $L$ common neighbors,  they are connected with a link definitely, as shown in Fig. 8 (a). $\delta=0$ means that the network is  totally unpredictable for the CN index. As shown in Fig. 8 (b),  any  node pairs of CN $>0$ are not connected.
Generally, $\delta$ is between 0 and 1 for most of the real-world networks, and it is dependent on $L$. If $\delta$ increases with $L$,  the node pair of a large CN value has large probability to be connected with a link, which means the network can be precisely predicted by the CN index.

\begin{figure}
 \centering
%\onefigure{F8.eps}
\includegraphics[width=5in,height=1.5in]{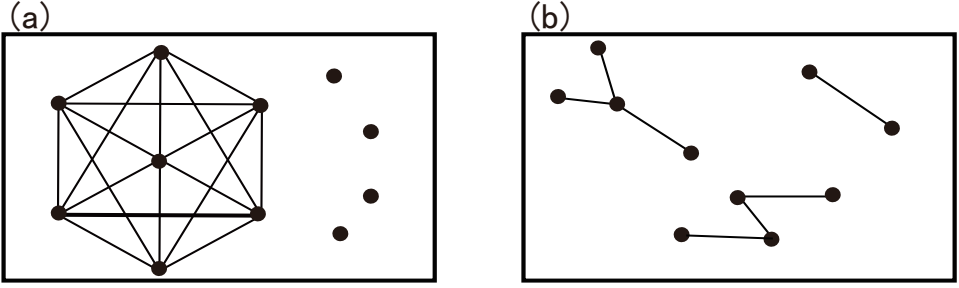}
\caption{Two extreme cases for illustration. In (a), if any two nodes  have common neighbors, they are connected by an edge definitely. Thus, the self-predictability $\delta$ of (a) equals 1. In (b), if any two nodes have common neighbors, they are not connected by an edge. The self-predictability  $\delta$ of (b) is 0. }
%\label{fig.8}
\end{figure}

\subsection{Simulation results}
We calculate the self-predictability of several large-scale real-world networks \cite{leskovec-ca-astroph,jmcauley-facebook-gplus,leskov-web-Google-web-stanford,Domenico-higgs-twitter}, of which the statistics are shown in Table 2. In the simulation, our fast algorithm is employed to generate the node pairs with CN values greater than $L$. In Fig. 9, we see that for most of the real-world networks, $\delta$ increases with $L$. This indicates that node pairs of large CN values have large probability to be connected with links, and these node pairs form the dense area of a network which embodies the intrinsic organization rule of a network. However, there are fluctuations and exceptions in the curves. For example, for Gplus, when $L$ increases from 5000, $\delta$ even decreases with $L$.  This indicates that besides the ``CN rule"  there are some other factors  affecting the organization and evolution of a network.

\begin{figure}
 \centering
%\onefigure{F5.eps}
\includegraphics[width=\textwidth,height=3in]{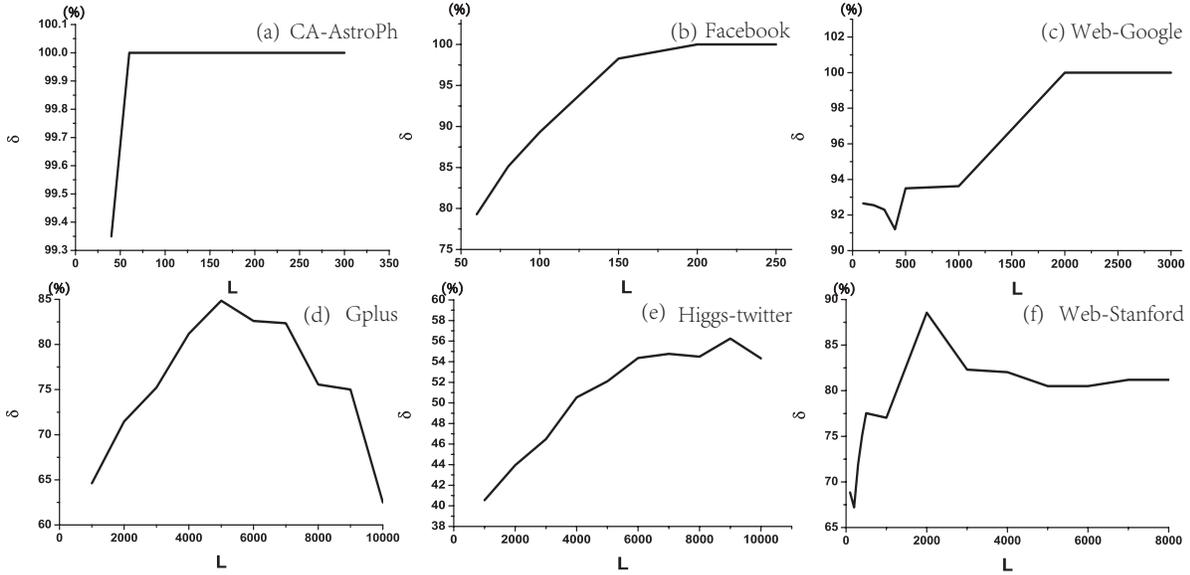}
\caption{Self-predictability $\delta$ vs lower bound $L$ for large-scale real-world networks. }
%\label{fig.5}
\end{figure}

\section{Conclusion and discussion}
Although CN-based indices for link prediction have  attracted much attention in the past few years, their performances are hard to evaluate in large and dense real-world networks.
In our framework, to efficiently and precisely predict the links, only node pairs of CN values greater than the lower bound are considered. This is mainly based on the fact that in real society hot individuals are more attractive and are more likely to be recommended than the others.
Then, we present two lemmas, based on which we further propose a fast parallel algorithm to calculate the CN values.
 Our algorithm works much more efficient than the other related algorithms in that by two delicate filtering operations, it greatly excludes the node pairs of CN values no greater than the lower bound in advance of the CN calculation. Thus, our algorithm is especially applicable for large-scale  real-world networks, since these networks usually have heterogeneous CN value distributions, where a large number of node pairs just have  small CN values, while a small fraction of node pairs have large CN values. Moreover, the efficiency of our algorithm increases exponentially with the increase of the lower bound of CN values.

Then, we propose a new metric, which is the self-predictability, for evaluating the performance of a similarity index in  link prediction. Calculation of self-predictability  needs not to divide the network into the train and probe sets, and thus it requires a lower computational cost than the metrics such as precision and AUC.

We employ our fast algorithm to calculate the self-predictability of many large-scale real-world networks. We find that generally self-predictability increases with the lower bound of  CN values, which indicates that two nodes with more common neighbors are more likely connected by a link. On the other hand, we find that there are fluctuations and exceptions in the simulation results of self-predictability, which reflect that  the ``CN rule" is not the only law that governs the organization and evolution of a network.

It is worth remarking that besides link prediction our fast algorithm can be also applied to the other CN-based problems in very large real-world networks. Also, the self-predictability is discussed in the context of CN index, while it can be easily generalized to evaluate the performance of the other similarity indices in link prediction.

\appendix
\section{The pseudocodes of two traditional MapReduce-based algorithms }
The pseudocode of the pair generating algorithm \cite{papadimitriou} is shown in Table A1. The pseudocode of the vectorization algorithm \cite{sschelter} is shown in Table A2.

\begin{table}
\centering
\begin{tabular}{p{\textwidth}}
%\hline \textbf{Algorithm 2}: \\
\hline map($u$, adj$\_$list[$u$] ) $\lbrace$ \\
\quad \textbf{foreach} $x$ in adj$\_$list[$u$]   \\
\qquad \textbf{foreach} $y$ in adj$\_$list[$u$]  \\
\qquad \quad if$(x<y)$  \qquad //to eliminate the repetition  \\
\qquad \quad \quad emit(relation$=(x,y)$,promote$\_$score$=1$) \\
$\rbrace$  \\
reduce(pairList) $\lbrace$ \\
\quad for pair in pairList \\
\quad \qquad add one score to the pair \\
$\rbrace$ \\ \hline
\end{tabular}
\caption{The pair generating algorithm.}
\end{table}

\begin{table}
\centering
\begin{tabular}{p{\textwidth}}
%\hline \textbf{Algorithm 3}: \\
\hline map($u$, adj$\_$list[$u$] ) $\lbrace$ \\
\quad \textbf{foreach} $x$ in adj$\_$list[$u$]   \\
\qquad subarray$=$the nodes which number smaller than $x$  \\
\qquad emit($x$,subarray)to reduce process  \\
$\rbrace$  \\
reduce($x$,[ subarray1,aubarray2,$\cdots$ ]) $\lbrace$ \\
\quad \textbf{foreach} arrayItem in [ subarray1,aubarray2,$\cdots$ ] \\
\quad \qquad \textbf{foreach} $y$ in arrayItem \\
\quad \qquad \qquad add one score to the pair$(x,y)$ \\
$\rbrace$ \\ \hline
\end{tabular}
\caption{The  vectorization algorithm.}
\end{table}
\section{Implementation of our  algorithm based on RDD}
The resilient distributed datasets (RDDs) is an efficient, general-purpose and fault-tolerant abstraction for sharing data in cluster applications. RDDs can efficiently express many cluster programming models including MapReduce, SQL, Pregel and so on. Here, we also express our algorithm with RDD, which can be further implemented in Spark. On the other hand, in  shared memory environment we can randomly access the data with its address, which can further reduces the amount of data emitted from the map function to the reduce function. The RDD description of our algorithm for shared memory environment is shown in Table B1. With RDD, the performances of our algorithm such as the volume of communication traffic, space utilization,  etc. can be further optimized.
\begin{table}
\centering
\begin{tabular}{p{\textwidth}}
%\hline \textbf{Algorithm 5}:Implementation of algorithmic framework based on random access environment \\
\hline val filtered $=$ adjGraph.filter($s =>$ $s$.$\_$$2$.size$>L$)  \qquad //step 1 \\
\quad .flatMap($s =>$ for (i $<-$ $s$.$\_$$2$) yield (i, $s$.$\_$$1$) ).groupByKey() \\
\qquad .zipWithIndex()  \qquad// step2  \\
val result $=$ filtered.flatMap( $s =>$  \\
\quad for (i $<- 1$ until $s$.$\_$$1$.$\_$$2$.size) \\
\qquad yield($s$.$\_$$1$.$\_$$2$(i), ($s$.$\_$$2$, i) ).groupByKey()  \qquad // step3 \\
\quad .filter($s=>s$.$\_$$2$.size$>L$).flatMap $\lbrace$ $s$ $=>$\\
\qquad for (lnk $<-$ $s$.$\_$$2$.$\_$$2$; i $<-$ $0$ until lnk.$\_$$2$) \\
\qquad \quad yield (($s$.$\_$$1$, getByIndex(filtered, link.$\_$$1$).$\_$$1$.$\_$$2$(i)), $1$)\\
\quad $\rbrace$.reduceByKey($\_$$+$$\_$).filter($s=>$$s$.$\_$$2$$>L$)  \qquad // step4 \\
 \hline
\end{tabular}
\caption{The pseudocode of our algorithm with RDD.}
\end{table}

\end{document}